\documentclass[twocolumn,english,prb]{revtex4}
\usepackage[T1]{fontenc}
\usepackage[latin9]{inputenc}
\usepackage{graphicx}

\usepackage{amsfonts}

\makeatother
\usepackage{babel}

\begin{document}

\title{Quantum charge transport in Mo$_{6}$S$_{3}$I$_{6}$\ molecular
wire circuits}

\author{M. Uplaznik, B. Bercic, M. Remskar and D. Mihailovic}

\affiliation{Jozef Stefan Institute, Jamova 39, SI-1000,Ljubljana, Slovenia}

\begin{abstract}
Charge transport measurements on flexible Mo$_{6}$S$_{3}$I$_{6}$\ (MoSI) nanowires
with different diameters in highly imperfect 2-terminal circuits reveal systematic power
law behaviour of the conductivity $\sigma(T,V)$ as a function of
temperature and voltage. 
On the basis of measurements on a number of circuits we conclude that the behaviour in \emph{thin}
wires can be most convincingly described by tunneling through Tomonaga-Luttinger
liquid (TLL) segments of MoSI wire, which is in some cases modified by environmental Coulomb blockade (ECB). 
The latter are proposed to arise from deformations or imperfections of the MoSI wires, which - in combination with their recognitive
terminal sulfur-based connectivity properties - might be useful for creating sub-nanometer scale interconnects as well as non-linear elements for molecular electronics. 
\end{abstract}
\maketitle

\section{Introduction}

While the transport properties of one-dimensional systems have been
of great interest from the point of view of fundamental physics for
some time\cite{TLL}, recently, further interest in the transport
properties of nano-scale one-dimensional systems was aroused because
of their importance for the development of molecular electronics,
where diverse molecular devices (switches, memory elements, sensors)
all need to be self-assembled together with electrically conducting
molecular-scale wires. To be of practical use, the connectors need to have
reliable contacts and also be able to withstand mechanical deformations
while retaining their conducting properties. Till now there has been
no recognized material which could be used for this purpose, and this
has seriously impeded progress in the development of large scale molecular electronics in recent
years.

In this paper we investigate the transport properties of Mo$_{6}$S$_{3}$I$_{6}$
molecular wires\cite{Mihailovic}, which have been recently shown to be very promising
flexible molecular-scale conductors. Mo$_{6}$S$_{3}$I$_{6}$ wires
are air-stable one-dimensional inorganic cluster polymers (see Fig.1),
which are unique in that they enable covalent bonding to gold surfaces
and organic molecules via sulphur atoms at the ends
of each molecular wire\cite{Ploscaru}. Single molecular wires were also recently shown to self-assemble gold particles into critical scale-free networks \cite{Strle}. Such molecular wires may expect their electron
transport properties to be governed by quantum properties on the microscopic
level. Thus, to make further progress in molecular electronics with
MoSI connectors, we must first investigate and understand their molecular
scale electronic transport properties. Earlier experimental work
has shown metallic signatures, such as a low-frequency Drude response
in the optical conductivity\cite{Vengust}, but relatively low room
temperature conductivities $\sigma\simeq10$ S/m,
which decrease with decreasing temperature\cite{Uplaznik}. Recently Venkataraman,
Hong and Kim \cite{Kim} described electron transport measurements
in multichannel Li$_{2}$Mo$_{6}$Se$_{6}$ nanowires with diameters
in the range 7.2 to 12 nm in terms of a Tomonaga-Luttinger liquid
(TLL) in contact with Fermi liquid (FL) electrodes. In another related material
NbSe$_{3}$, nanowires ranging between 30 and 300 nm in diameter have been observed to display similar power-law
behaviour, albeit over a rather limited range of temperature\cite{Slot}. The
extreme one-dimensional nature of the MoSI wires suggests that signatures of TLL
behaviour might also be observed in thick multi-strand bundles, and not only in very thin wires.
On the other hand, we may expect that the electronic transport properties in
MoSI wires might depart from ideal TLL behaviour because of the deformable
nature of the S bridges which link together the Mo$_{6}$ clusters
into 1D chains \cite{Vilfan}. 

\begin{figure}[h]
\begin{centering}
\includegraphics[width=8cm,height=8cm,keepaspectratio]{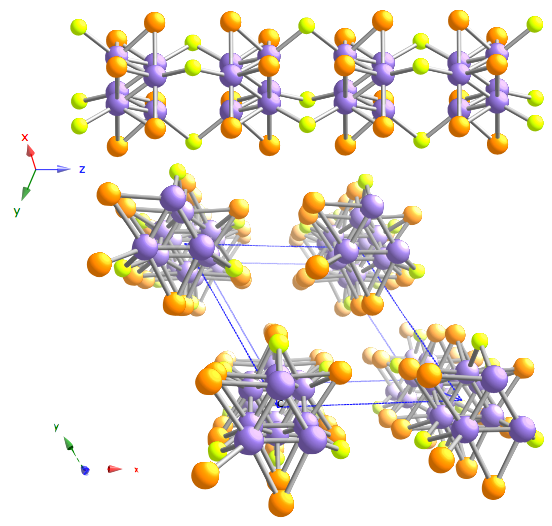} 
\par\end{centering}
\caption{The Structure of Mo$_{6}$S$_{3}$I$_{6}$ wires. The axes are in
the frame of the monoclinic space group P$\bar{1}$ \cite{Nicolosistructure} with center-to-center distance of $d_0=0.958$, and the unit cell length $c= 1.197$ nm. The S atoms linking
the Mo clusters form accordion-like deformable bridges and are shown in yellow, while the orange and violet spheres
represent I and Mo respectively. }
\end{figure}

In this paper we present a systematic study of MoSI wires of different
diameter from 4 nm to 1000 nm, examining the $T$-dependence, diameter
dependence and current-voltage characteristics at different temperatures.
We are particularly interested in the behaviour of realistic circuit
configurations, with irregular wire geometries. The intrinsic flexibility
of the MoSI wires, arising from their accordion-like structure \cite{Meden,Nicolosistructure}
means that they bend easily to conform to surface contours. We have
therefore focused on dielectrophoretically deposited thin wires over
contacts in which the wires conform to the surface relief. 

Considering the possible transport mechanisms, we confine ourselves
to the common ones discussed in literature\cite{Kim,Slot}, namely TLL tunneling\cite{TLL}, environmental Coulomb blockade
\cite{ECB} (ECB) and variable range hopping (VRH) in the presence of Coulomb charging effects \cite{VRH}:

1. The $T-$ and $V-$ dependence for tunneling into a 1D TL
liquid via Fermi-liquid metal contacts is given by: 
\begin{equation}\label{eq_LL_IV_final}
    I=I_{0}T^{1+\alpha}\sinh\left(\frac{\gamma eV}{2kT}\right)
    \left|\Gamma\left(1+\frac{\beta}{2}+i\frac{\gamma eV}{2\pi kT}\right)\right|^{2}
\end{equation} 

where $\alpha = (g^{-1}-1)/4,  \beta = (g+g^{-1}-2)/8$ 
 and the Luttinger parameter $g=v_{F}/v_{\rho}$.  $\gamma$ is a fitting parameter that accounts for the
voltage drop over the circuit \cite{Bockrath,Kim}.
     A collapsed
diagram of the underlying transport characteristic is obtained by plotting $I/T^{\alpha+1}$
against $eV/kT$, where $\alpha$ is the slope of zero voltage conductivity
against temperature $\sigma=\sigma_0T^\alpha$. $\beta$ is the exponent for the
high voltage limit ($eV\gg kT$) arising from the 
power law behavior $I\propto V^{\beta+1}$.

2. Unfortunately ECB models cannot
be solved analytically for the general case, but the asymptotic behavior is very characteristic. The experimentally obtainable low-temperature behaviour is given by:
 \begin{equation}\label{eqECBT_I(V)_final_ohmic}
    \sigma_{T\rightarrow0,V\rightarrow0}=\left(\frac{2}{g}+1\right)\frac{e^{-2\gamma/g}}{\gamma(2+2/g)}\frac{1}{R_{T}}
    \left[\frac{\pi}{g}\frac{e|V|}{E_{C}}\right]^{\frac{2}{g}}.
\end{equation} where $E_{C}$ is the charging energy, $g=G_{0}/G$, \emph{G}
is the frequency-independent conductance, $G_{0}=2e^{2}/h$ and $R_{T}$ is the tunneling resistance.\cite{ECB}
For low voltages and temperatures the current follows a power
law behavior $I\propto V^{2/g}$. For
high voltages (but low temperatures), \begin{equation}\label{eqECBT_I(V)_highE}
    I(V)=\frac{1}{R_{T}}\left[V-\frac{e}{2C}+\frac{g}{\pi^{2}}\frac{e^{2}}{4C^{2}}\frac{1}{V}\right]
    \qquad
\end{equation} 
which gives a linear $I-V$ dependence at high V, so the derivative $dI/dV$ is expected to approach an asymptotic constant value of $1/R_{T}$. \cite{ECB,Uplaznik} It also gives a non-zero intercept for $I=0$ given by the charging energy $e^{2}/2C$.

3. For the variable range hopping mode, a plot of $\ln(G)$ (for low voltage) against $T^{-\lambda}$
yields curves which become linear with the correct hopping exponent
$\lambda$. The fits to the data typically give a large error in the exponents, so to extract the best value of $\lambda$, fits to the data are
tested statistically by calculating Pearson's correlation coefficients.

Rather than choosing a few measurements which obey one or another
type of behaviour, we present here a summary of a number of
experiments, to try and understand the different types of behaviour
that can arise in nanoscale circuits with MoSI wires.

\begin{figure}
\begin{centering}
\includegraphics[width=8.5cm]{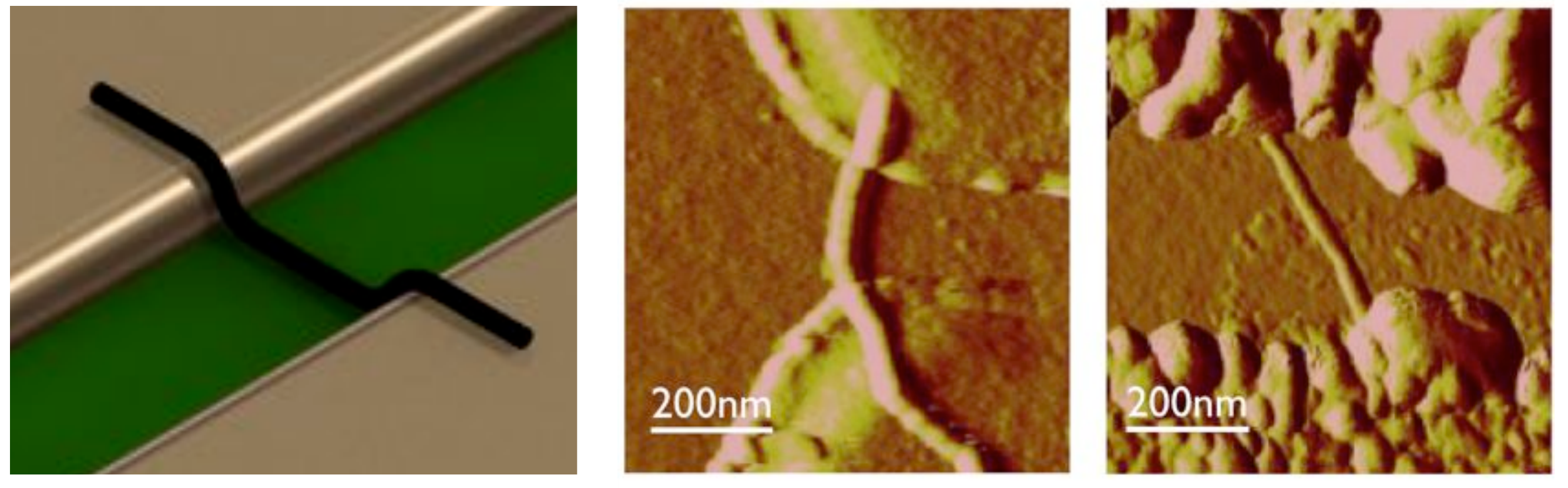} 
\par\end{centering}
\caption{A schematic diagram of the circuit geometry (left). AFM topographic images of the thin wires after annealing. The wire
diameters (from the height profile) are 4.2 and 4 nm for samples na23 (middle)
and na27 (right) respectively. The roughness of the Ni electrodes appears after
annealing and is caused by metal aggregation during the high temperature
annealing process.}
\end{figure}

\section{Experimental details}

The thin wires were prepared according to the method reported by Nicolosi
et al \citet{Nicolosi} by repeated dispersion and dilution. The dispersion procedure separates the wires into two distinct categories: thin wires, with
diameters $4<D<10$ nm and thick multichannel bundles with $100<D<1000$
nm\citet{Nicolosi}. In the ultrasonic bath processing procedure, the defective wires break up into shorter
segments, leaving less defective long thin wires
in solution. The individual strands within the thin
wires may thus be expected to have significantly fewer imperfections than
within the thick bundles, which may result in different electron transport behaviour in thick and thin wires. This should be evident in the room
temperature conductivity ($\sigma_{300K}$) as well as the $T$-dependence
and $I-V$ systematics.

The wires were dielectrically deposited typically over nickel electrodes prepared by electron beam lithography
(EBL) by placing a drop of solution over the electrodes and applying
a 50Hz AC electric field to the electrodes. The entire circuits were then annealed in
vacuum at 700 C for an hour. The circuits after annealing are shown
in Figure 2. Typical resistances of the nanowires at room
temperature were between 100k$\Omega$ and 100M$\Omega$. Care
was taken to ensure good thermal contact of the sample with the cryostat
cold finger.

\begin{figure}
\begin{centering}
\includegraphics[width=8cm]{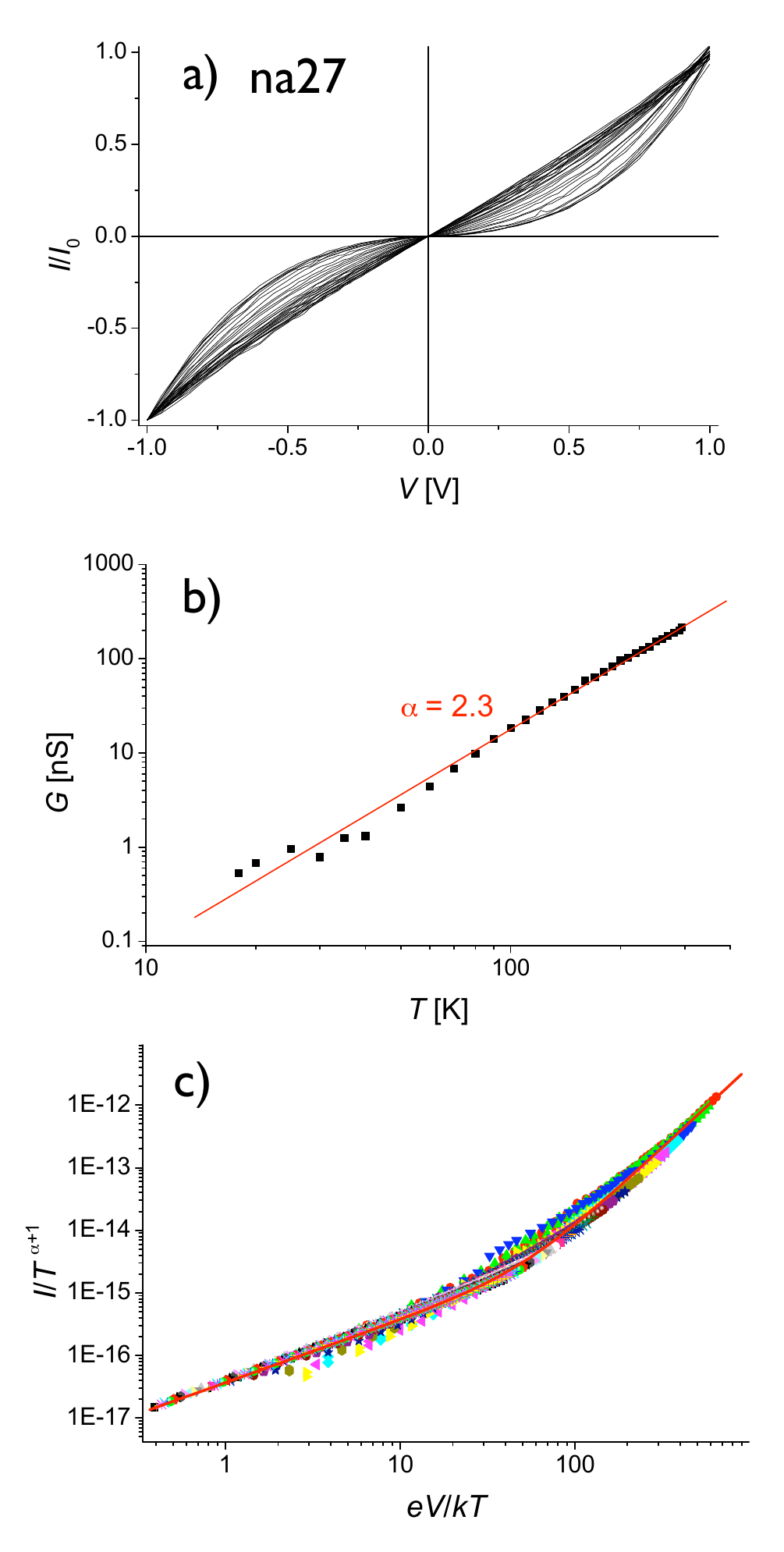} 
\par\end{centering}
\caption{a) Normalised $I-V$ characteristics of sample circuit labeled na27  ($d=4$nm) at different
temperatures, where $I_0=230$ nA. b) The conductance as a function of temperature on a
log-log plot. c) A plot of $I/T^{\alpha+1}$ vs. $eV/kT$ shows the data collapse onto a single curve. }
\end{figure}

\begin{figure}
\begin{centering}
\includegraphics[width=8cm]{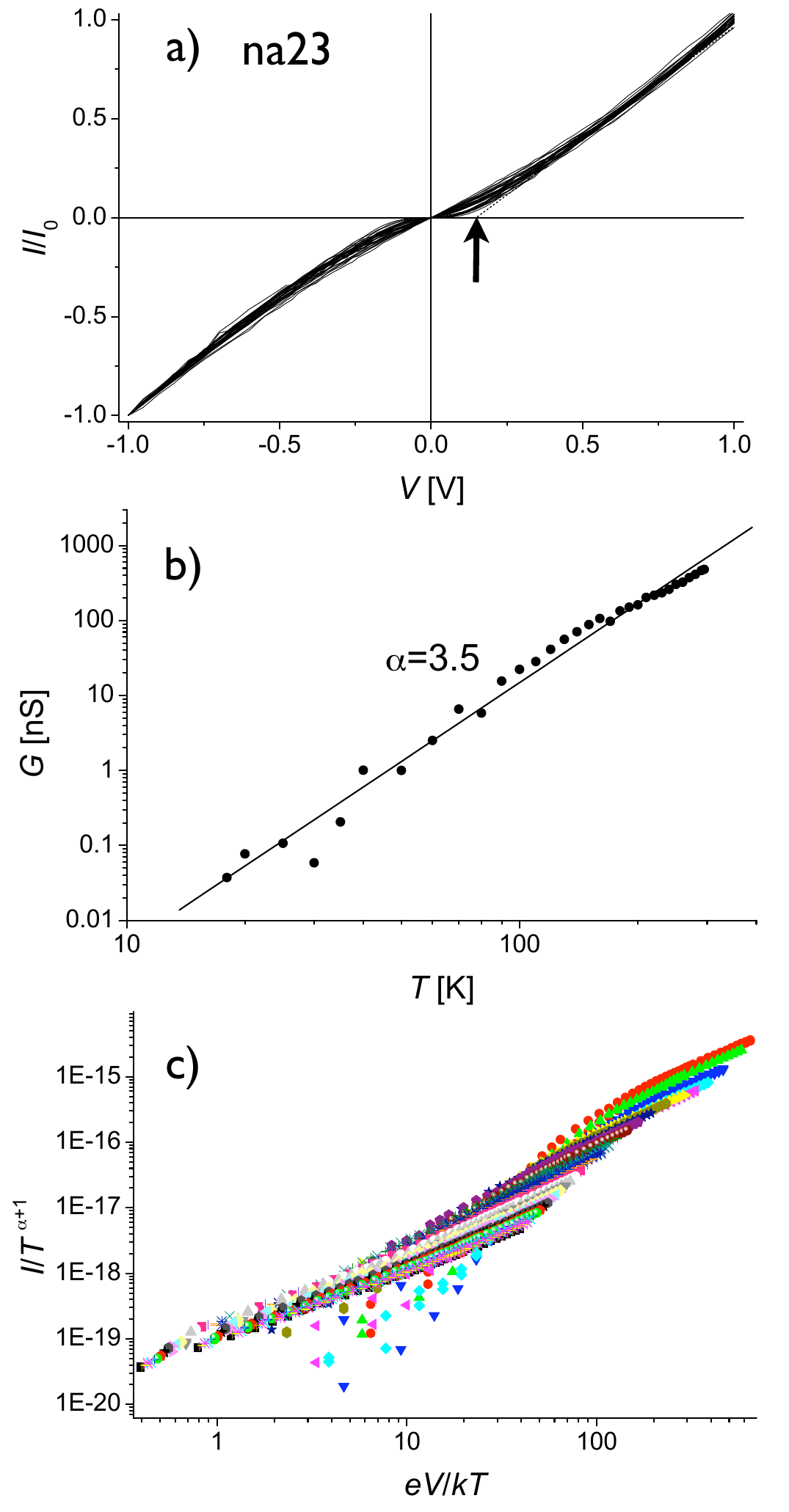} 
\par\end{centering}
\caption{a) Normalised $I-V$ characteristics of sample circuit labeled na23  ($d=4.2$nm) at different
temperatures, where $I_0=600$ nA. The arrow shows the extrapolated voltage corresponding to $V=e/2C$. b) The conductance as a function of temperature on a
log-log plot. c) A plot of $I/T^{\alpha+1}$ vs. $eV/kT$ do $not$ show the data collapse exhibited by circuit na27 in Fig. 3c). }
\end{figure}

\begin{figure}
\begin{centering}
\includegraphics[width=8cm]{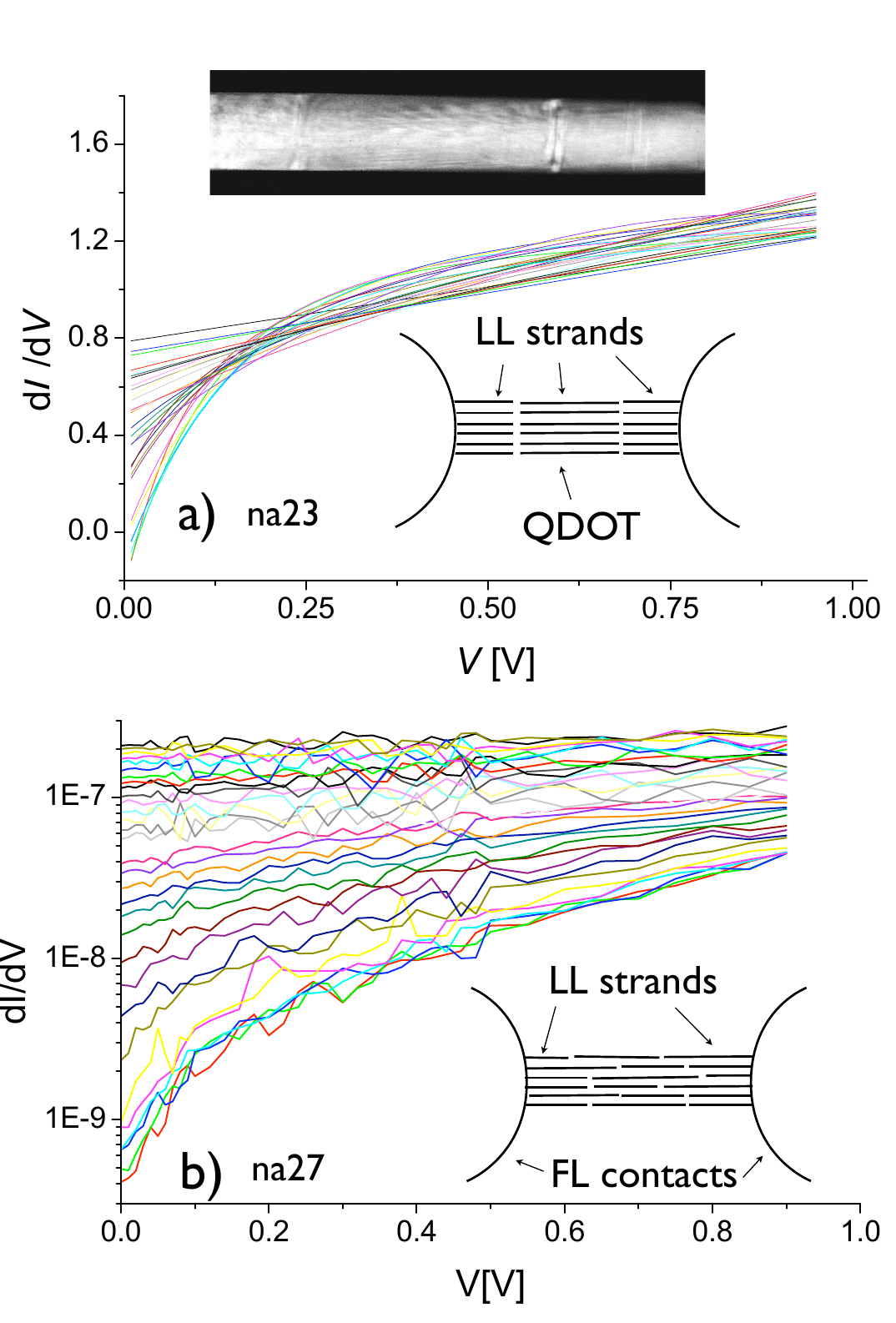} 
\par\end{centering}
\caption{The derivatives $dI/dV$ obtained from the I-V data plotted in Figs. 3a) and 4a) for circuits a) na23 and b) na27 at different
temperatures. The derivative at low $V$ increases in both cases, as T increases from 18K to 300K }
\end{figure}

\begin{figure}
\begin{centering}
\includegraphics[width=8cm]{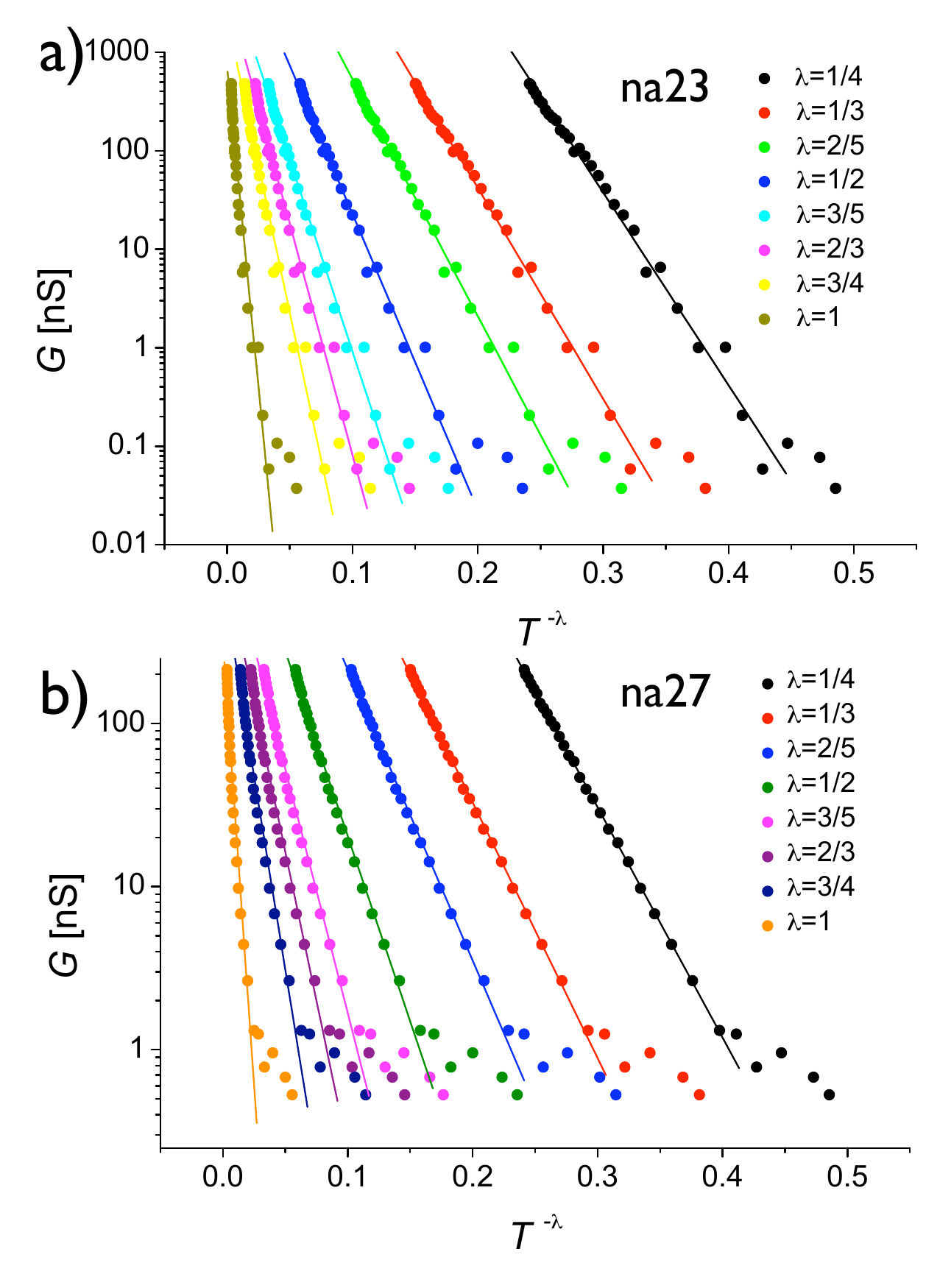} 
\par\end{centering}

\caption{I-V characteristic of sample circuits a) na23 and b) na27 at different
temperatures. }

\end{figure}

\section{Experimental results and analysis}

\subsection*{Thin wires}

Two thin wire circuits labeled na27 and
na23 (shown in Fig. 2) illustrate the most common type of behaviour
observed in a number of measured circuits with diameters ranging from $d\sim4$ to
15 nm. The room-temperature conductivity $\sigma_{300K}$ for na27 and
na23  was 3710 S/m and 11900 S/m respectively. Their $I-V$ characteristics
at temperatures between 18 K and 300K are shown in Figs.3a) and
4a) and exhibit qualitatively different behaviour. In the
case of sample na27, the I-V curves show characteristic inverted S-shaped
curves whose curvature is strongly T-dependent. In  contrast, circuit
na23 shows a clear J-like shape characteristic. (Other circuits
we have measured show behaviour in between these two extremes \cite{MUThesis}). The
$T$-dependence of the conductance $G=dI/dV$ at $V=0$ is shown in Figs. 3b) and 4b). The line is a fit to a power law $G=G_0 T^{\alpha}$ where $\alpha=2.3$
and $3.5$ respectively. The data follow the power law fit reasonably
well, but do not give a perfect fit over the entire range of T. Plotting
the entire data set $I/T^{\alpha+1}$ against $eV/kT$ according to
the TLL prediction, for na27 we see that the data collapse quite well
onto a single curve (Fig. 3c), where $\beta=1.6$
is obtained from the fit. The behaviour of na23 is very different to na27. It gives no such TLL collapse (Fig. 4c), indicating clear departure
from TLL predictions. (Overall, approximately half our circuits showed
the TLL collapse.)

Attempting to fit the ECB
model to the data, we would expect the $I/V$ slope to cross over from $\sim2/g$ at
low voltages to $1/R_T$ at high voltages. Correspondingly, the derivative $dI/dV$
should show a systematic $T$-independent
cross-over from $2/g-1$ to 0 corresponding to the low and high
voltage limits respectively. We plot $dI/dV$ vs. $V$,  in Figs. 5 a) and b) respectively for na23 and na27. 
Circuit na27 shows a rather small variation of $dI/dV$ with V over the entire range of $V$ and $T$ (note the small values on the log scale on the $dI/dV$ axis), quite unlike the ECB model predictions. In contrast, na23 shows a systematic variation with $V$, which for low temperatures is not far from what is expected on the basis of ECB theory: at low $V$, $dI/dV \sim 0$ (within experimental error), but clearly saturating at high $V$.

For completeness, attempting to fit the data to the VRH model, 
we first plotted $\ln(G)$ (for low voltage) against $T^{-\lambda}$
for different $\lambda$=1/4 to 1, according to the VRH model (Fig.6). Making a statistical evaluation using
Pearson's correlation test, we find that the best fit is with $\lambda=1/4$
for na27 and $\lambda=1/3$ for na23. Such small values of $\lambda$
within VRH models would imply the dominance of 3D hopping processes, which is not expected.
(We shall discuss the implications later).

\subsection*{Thick bundles}

A ubiquitous feature of the thick wire bundles ($d=100\sim1000$)
is their linear I-V characteristic from 18K to 300K \cite{Uplaznik}.
The room temperature conductivity $\sigma_{0}$ is typically 3 orders
of magnitude \emph{smaller} than for the thin wires, around 1-10 S/m.
This is taken as a clear indication that only a small fraction $f$ of
the molecular strands in the bundle contribute to the transport. Confirming
earlier preliminary measurements, the T-dependence was not very well
described by a 1D VRH model\cite{Uplaznik}. The new systematic data
on many circuits now confirms this. The dependence of the conductivity
is shown in Figure 7 for a number of bundles of different diameter.
Surprisingly, the data for the different bundles all appear to follow
power-law behaviour $\sigma=\sigma_{0}T^{\alpha}$ quite well. Moreover,
a systematic trend is observed, whereby the larger diameters have
a smaller exponent $\alpha$, an indication that 
the numbers of conducting channels scales with diameter\cite{TLL} (Figure 8).

\begin{figure}[h]
\begin{centering}
\includegraphics[width=8cm,height=8cm]{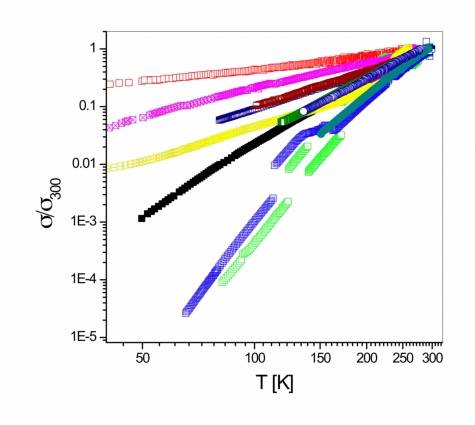} 
\par\end{centering}
\caption{The normalized conductance $G/G_0$ of a large number of thick wires, with $d=200 \sim2000$ nm. In two measurements, a break is obseved, which is believed to be strain-induced change in the number of conducting channels. }
\end{figure}

\begin{figure}[h]
\begin{centering}
\includegraphics[width=6cm,height=6cm]{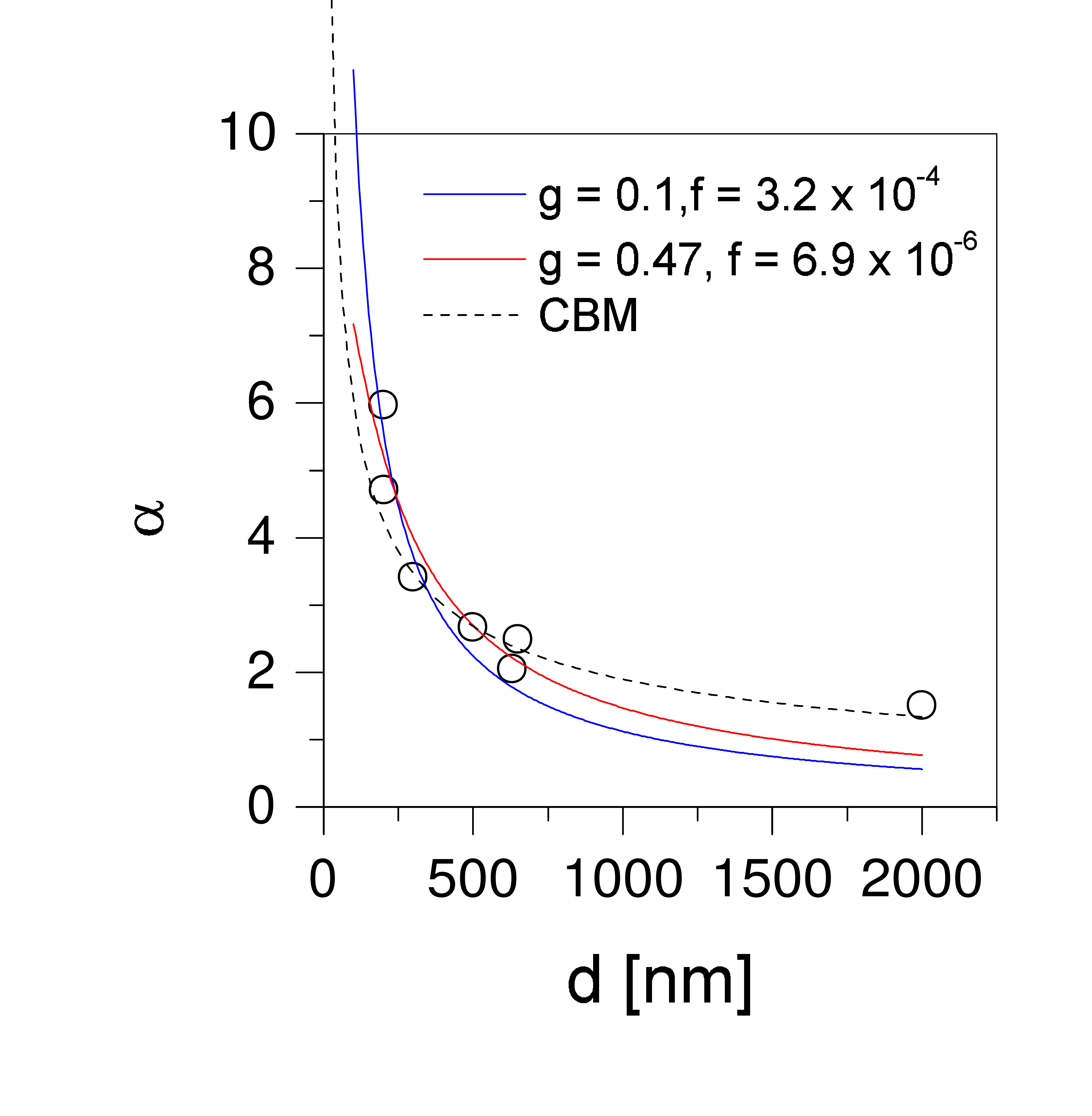} 
\par\end{centering}
\caption{The dependence of the temperature exponent $\alpha$ of thick wires on bundle diameter $d$ from the data in Fig. 7.}
\end{figure}

\section{Discussion}

The present set of experiments shows a rather wide range of behaviour,
suggesting caution in trying to interpret the data by any single model. 
Discussing the simple VRH model first, our fits give 3D exponents, which is clearly inconsistent with the nature of the system investigated here and is dismissed from further discussion. The calculation of Fogler et al. \cite{VRH}
considered a quasi-1D wire with a finite density of impurities modeled by a series of weakly coupled quantum dots.
They predict power law behaviour of the current for both T and V,
as $I\propto V^{\beta+1}$ at high $V$ and $I\propto T^{\alpha}V$
at low $V$ with $\alpha\gg\beta\gg1$. The model holds for a large
number of parallel statistically independent conduction channels,
where power laws are not obscured by the fluctuations of G. This case
is fulfilled here for thick wires, with bundle diameters between 100 nm and 1 $\mu$ might have
$10^{4}$ to $10^{6}$ conducting channels. However,
the prediction that $\beta\gg1$ is not fulfilled here for either
the thin or the thick multichannel bundles, suggesting that this model may not be applicable for describing MoSI circuits. 

In circuit na 27 the TLL prediction can be confirmed with some degree of confidence by the collapsed plot of $I/T^{\alpha+1}$ vs. $eV/kT$. For tunneling from a Fermi liquid into a perfect TLL without defects, the model prediction is that $\alpha=\beta$. Imperfections, such as deformations and kinks or stoichiometric defects break up the wire into TLL segments, which introduces TLL-TLL tunneling between these segments\cite{Kim}, for which the predicted ratio changes to $\alpha=2\beta$. The measured value  for na27 is $\alpha/\beta=1.45\pm0.1$. A similar circuit with a wire diameter of 5 nm (na12, not shown) gives a near-identical collapse as na27 with $\alpha/\beta=2.00\pm0.1$. 

The exponent $\alpha$ for a $N$-channel wire is given by TLL theory\cite{MG}
as:

\begin{equation}
\alpha=\frac{2}{N}[\sqrt{1+\frac{2N}{g^{2}}}-1],\label{1}
\end{equation}

where $g$ is the electron-electron interaction parameter. The number of conducting channels
$N$ in a nanowire of diameter $d$ can be written as $N=2f\frac{D^{2}}{a^{2}}$,
where $a$ is the lattice constant, and $f$ is the fraction of molecular
wires which actually carry current without interruption. For circuit na27, using $f=1$, $\alpha=2.3$, and $N=16$ channels (for $d= 4$ nm), we calculate $g=0.21$. This is slightly larger than the value 0.15 obtained by Venkataraman et al for Li$_{2}$Mo$_{6}$Se$_{6}$ nanowires. Other wires which we have measured have $\alpha$ between 2 and 3.5, giving a range of $0.09<g<0.27$.  
Comparing this with $g$ estimated from materials parameters, we can use the expression from the Coulomb charge screening model\cite{MG} $g^{2}= v_{F}R_{0}C_{0}$, where $v_{F}=7 \times 10^{5}$m/s is the Fermi velocity calculated by DFT calculations, $R_{0}$ is the resistance quantum and the single wire capacitance per unit length is $C_{0}=\frac{2\pi\epsilon_{r}\epsilon_{0}}{\text{ln}[4t/d_{0}]}$. Using $\epsilon_{r}=3.9$ for the Si oxide layer of thickness $t=600$nm, we obtain $g=0.49$. In this estimate, the DFT value of $v_{F}$ is likely to be overestimated, which may account for a large part of the discrepancy between the values obtained directly from the measurements. It has also been noted previously that the static Coulomb screening model overestimates $g$ \cite{Kim}.

The data in Figure 8 for the thick wires can be fit using the expression (4). A range of values of
$g$ can be obtained from the fits to the present data, $0.1<g<0.5$
with corresponding filling factors $10^{-5}<f<10^{-4}$. Values outside
this range of $g$ cannot be made to fit the data. However, since the $I-V$ characteristics are ubiquitously linear, for the thick wires, this in itself cannot be taken as proof for TLL behaviour.  

The low value of $\sigma_{0}$ in thick wires is assumed to be related to imperfections on the wires. It suggests
that the transport along the wires is dominated by tunneling between
TLL segments. In fact the resistance of both thick and thin wires
is typically of the same order of magnitude, which suggests that in the
thick bundles only a very small number of wires are uninterrupted, or that
only strands on the outside of the bundle are conducting, with the
inner strands being unreachable due to the extremely small perpendicular inter-molecular
hopping rates within each wire.

Turning to the case of na23, which shows clear characteristics of ECB behaviour, the intercept of the $I-V$ curve at $I=0$ gives $V=\frac{e}{2C}\simeq 0.15 $V and consequently $C\simeq 5.3\times 10^{-19}$F. Comparing this QDOT capacitance with an estimate of C for a single wire over a Si ground plane with a 600 nm SiO$_{2}$ insulating layer and length $L=265$ nm between contacts, we have $C\simeq 8 \times 10^{-18}$, for na23. Depending on what we assume for the QDOT shape, its size appears to be some fraction ($\sim 1/10$) of the distance between the electrodes. 

Considering the possible origin of the ECB behaviour and the observed departures from
the predicted TLL characteristics, we can imagine imperfections and breaks of continuity in the wires of diverse origin. The most obvious and unavoidable effect arises from the deformation of the nanowires adapting to the relief of the contacts (Figure 2). Their inherent flexibility allows for a substantial deformation, which is accompanied by significant changes in electronic structure near the Fermi energy \cite{Vilfan,Tomanek}. We can thus envisage that the deformations can lead to breaks in the continuity of individual channels, and as a result the formation of a QDOT in between the breaks. In this case the ECB capacitance would scale with the distance between electrodes. Another effect, which might be important arises from the intrinsic tendency of the wires to form discontinuities in the structure, as shown in the HRTEM image in Figure 5 a). Clear stripes are sometimes observed across the wire bundles, which arise from from stacking faults which appear during the growth process. Indeed such compositional ordering has been recently theoretically predicted\cite{Yang}.  HRTEM diffraction analysis shows that the structure of the wire is identical on both sides of the fault, but clearly continuity is broken at these points. Sections in between the faults may thus act as QDOTs, leading to the ECB behaviour we observe. Considering that the distance between faults may be a few tens of nanometers, the comparison of QDOT capacitance with the static capacitance calculated in the previous paragraph suggests this effect may also be important.

\section{Conclusion}

One of our objectives has been to determine how imperfect, bent and deformed MoSI wires might behave in molecular-scale circuits such as may form upon self-assembly\cite{Strle}. It is clear from the present experiments that quantum transport dominates their behaviour. The TLL model with its characteristic data collapse of the $I-V$ characteristics at different temperatures appears to hold well for a significant proportion of the thin wire circuits. At the same time clear signatures of the ECB predicted J-shaped $I-V$ curves are also occasionally observed. Discontinuities in the wires either as a result of bending, and/or structural stacking faults within the wire are believed to cause the formation of QDOTs, which lead to the occasional occurrence of ECB behaviour. The diameter of the present MoSI wire bundles is small enough to allow covalent S bonding to individual molecules\cite{Ploscaru}, so for the construction of molecular scale circuits, where thin and relatively short wires are of interest, they may potentially revolutionize molecular electronics. A point of interest is the possibility of making variable sizes of QDOTs with MoSI wires by stretching them over appropriately sized regular topological features to produce arrays of QDOTs. The other possibility of creating single QDOTs by using the tip of an atomic force microscope, was already recently demonstrated \cite{Abdou}.

We wish to thank D.Vengust for providing samples of MoSI nanowires and J.Strle for proof reading the manuscript.


\begin{thebibliography}{20}
\bibitem{TLL} J. Voit, Reports on Progress in Physics \textbf{58},
977 (1995) and references within.

\bibitem{Mihailovic}D.Vrbanic et al., Nanotechnology \textbf{15}, 635 (2004), for a review, see: D.Mihailovic, Rep. Materials Science \textbf{54}, 309 (2009)

\bibitem{Ploscaru} I.M. Ploscaru, S. Jenko Kokalj, M. Uplaznik, D.
Vengust, D. Turk, A. Mrzel, D. Mihailovic, Nanoletters, \textbf{7},
1445 (2007).

\bibitem{Strle} J.Strle, D.Vengust and D.Mihailovic, Nano Letters \textbf{9}, 1091 (2009)

\bibitem{Vengust} D. Vengust, F. Pfuner, L. Degiorgi, I. Vilfan,
V. Nicolosi, J.N. Coleman, D. Mihailovic, D.D., Physical Review. B:
Condensed Matter, \textbf{76}, 075106 (2007)

\bibitem{Uplaznik} M. Uplaznik, B. Bercic, J. Strle, M.I. Ploscaru,
D. Dvorsek, P. Kusar, M. Devetak, D. Vengust, B. Podobnik, D. Mihailovic,
Nanotechnology, \textbf{17}, 5142 (2006), B.Bercic et al Appl. Phys. Lett. \textbf{88} 173103 (2006)

\bibitem{Kim} L. Venkataraman, Yeon Suk Hong, P.Kim, Phys. Rev. Lett.,
\textbf{82}, 4918 (1999).

\bibitem{Slot} E. Slot, M.A. Holst, H.S.J. van der Zant, S.V. Zaitsev-Zlotov,
Physical Review Letters, \textbf{93}, 176602 (2004).

\bibitem{Vilfan} I. Vilfan, D. Mihailovic, Physical Review B, \textbf{74}
235411 (2006)

\bibitem{Nicolosistructure} Nicolosi et al., Adv Mater  \textbf{19}, 543 (2007)

\bibitem{Meden}T.Meden et al, Nanotechnology \textbf{16}, 1578 (2005)

\bibitem{Bockrath} M.C. Bockrath, H. David, J. Lu, A. Rinzler, R.E.
Smalley, L. Balents, P/I/ McEuen, Phys. Rev. Lett., Nature, \textbf{397},
596 (1999).

\bibitem{ECB} H. Grabert, M.H. Devoret, NATO ASI Series B, 294, (1992)

\bibitem{Nicolosi} V. Nicolosi, D. Vrbanic, A. Mrzel, J. McCauley,
S O'Flaherty, C. McGuinness, G. Compagnini, D. Mihailovic, W.J. Blau,
J.N. Coleman, The Journal of Physical Chemistry. B, \textbf{109} 7124
(2005)

\bibitem{VRH} M.M Fogler, S. Teber, B.I. Shklovskii, Physical Review.
B, \textbf{35} 035413 (2004)

\bibitem{MG} K.Matveev and L.Glazman, Phys.Rev.Lett. \textbf{70}, 990, (1993)

\bibitem{MUThesis} M.Uplaznik, PhD Thesis, Univ. of Ljubljana (2009)

\bibitem{Abdou} A.Hassanien et al., Physica E \textbf{29}, 684 (2005), H.W.Ch.Postma et al, \textit{Science} \textbf{293}, 76, (2001)

\bibitem{Tomanek}Popov et al.  Phys. Rev. Lett. (2007) \textbf{99} 085503 (2007)
\bibitem{Yang} T.Yang, S.Berber and D.Tomanek, Phys. Rev. B \textbf{77},7 (2008)

\end{thebibliography}
\end{document}